\documentclass[aps,prd,reprint,groupedaddress]{revtex4-2}

\usepackage[colorlinks,breaklinks=true,plainpages=false,citecolor=blue,linkcolor=blue,urlcolor=blue,bookmarksopen=true,bookmarksnumbered=false,bookmarksdepth=5]{hyperref}

\usepackage{mathtools}
\usepackage{lmodern}
\usepackage[T1]{fontenc}
\usepackage{bbm}
\DeclareMathAlphabet{\mathbfi}{OML}{cmm}{b}{it}

\let\originalleft\left
\let\originalright\right
\renewcommand{\left}{\mathopen{}\mathclose\bgroup\originalleft}
\renewcommand{\right}{\aftergroup\egroup\originalright}

\makeatletter
\newenvironment{equations}[1][]{\subequations\ifx\relax#1\relax\else\label{#1}\fi\align\ignorespaces}{\endalign\ignorespacesafterend\endsubequations}
\def\@spliteq#1{\begin{equation}\begin{split}#1\end{split}\end{equation}}
\def\splitequation{\collect@body\@spliteq}

\makeatother

\renewcommand{\vec}[1]{{\ifnum9<1#1\mathbf{#1}\else\ifcat\noexpand#1\relax\boldsymbol{#1}\else\mathbfi{#1}\fi\fi}}
\newcommand{\mathe}{\mathrm{e}}
\newcommand{\mathi}{\mathrm{i}}
\newcommand{\total}{\mathop{}\!\mathrm{d}}

\newcommand{\abs}[1]{{\left\lvert{#1}\right\rvert}}
\newcommand{\sgn}{\operatorname{sgn}}

\newcommand{\1}{\mathbbm{1}}
\newcommand{\eqend}[1]{\,#1}
\newcommand{\bigo}[1]{\mathcal{O}\left({#1}\right)}

\newcommand{\expect}[1]{\left\langle{#1}\right\rangle}

\allowdisplaybreaks

\begin{document}

\title{Noncommutative geometry from perturbative quantum gravity}

\author{Markus B. Fr\"ob}
\email{mfroeb@itp.uni-leipzig.de}
\affiliation{Institut f\"ur Theoretische Physik, Universit\"at Leipzig, Br{\"u}derstra{\ss}e 16, 04103 Leipzig, Germany}

\author{Albert Much}
\email{much@itp.uni-leipzig.de}
\affiliation{Institut f\"ur Theoretische Physik, Universit\"at Leipzig, Br{\"u}derstra{\ss}e 16, 04103 Leipzig, Germany}

\author{Kyriakos Papadopoulos}
\email[Corresponding author: ]{kyriakos@sci.kuniv.edu.kw}
\affiliation{Department of Mathematics, Kuwait University, Safat 13060, Kuwait}

\date{February 24, 2023}

\begin{abstract}
Trying to connect a fundamentally noncommutative spacetime with the conservative perturbative approach to quantum gravity, we are led to the natural question: are noncommutative geometrical effects already present in the regime where perturbative quantum gravity provides a predictive framework? Moreover, is it necessary to introduce noncommutativity by hand, or does it arise through quantum-gravitational effects? We show that the first question can be answered in the affirmative, and the second one in the negative: perturbative quantum gravity predicts noncommutativity at the Planck scale, once one clarifies the structure of observables in the quantum theory.
\end{abstract}

\maketitle

\section{Introduction}
\label{sec:intro}

Many competing theories exist that purport to quantize General Relativity (GR) or include a sector where gravity is quantized, such as string theory, loop quantum gravity, spin foams, geometrodynamics and the Wheeler--DeWitt equation, supergravity theories, noncommutative geometry, the asymptotic safety program, and many others. Conceptually, the quantization of GR is either done on the level of the gravitational part, quantising the field that carries the gravitational force (gravitons/metric fluctuations or their generalizations), or on the level of the geometrical part, quantising the underlying spacetime. However, since GR describes gravity as the curvature of spacetime, these different aspects are two sides of the same medal.

It is well-known that the quantization of gravity as a quantum field theory of metric fluctuations around a given background leads to a framework that is power-counting nonrenormalizable. This limits the predictive power of the theory, since new couplings need to be fixed experimentally at each order in perturbation theory, and thus in principle infinitely many. Nevertheless, it is viable as an effective field theory~\cite{burgess2003}, where one considers the theory at scales much less than a fundamental scale, such that the contribution of higher orders is suppressed with respect to lower orders, and can be neglected. This approach, known as perturbative quantum gravity, is a very conservative approach to a theory of quantum gravity since it results from the application of the well-established methods of Lagrangian quantum field theory to the Einstein--Hilbert Lagrangian describing classical gravity. For this reason, we expect that any theory of quantum gravity has to reproduce the predictions of perturbative quantum gravity in its region of validity, just as Newtonian gravity reproduces the results of GR in the weak-gravity regime.

On the geometrical side, a prominent approach to quantization is given by noncommutative geometry. In this framework, there are again many different approaches, for example Connes' noncommutative spectral triples~\cite{connes1995}, Lorentzian spectral triples~\cite{paschkeverch2004,franco2014}, Snyder and $\kappa$-Minkowski spacetimes and their curved-space generalizations~\cite{snyder1947,lukierskinowickiruegg1992,mignemi2010,ballesterosetal2019,franchinovinasmignemi2020}, or strict deformation quantization~\cite{grosselechner2007,buchholzlechnersummers2011,much2012,much2017}.\footnote{The literature on the topic is vast, such that we could only give some examples, and we refer also to the review~\cite{douglasnekrasov2001} and references in all these works.} The overarching general idea is to obtain classical geometry from the limit of a noncommutative algebra, which is conceptually analogous to the quantization of the classical phase space in the quantum-mechanical setting. A particular realization of noncommutative geometry, that is a quantum spacetime, is obtained by promoting the coordinates $x^\mu$ to noncommuting operators $\hat{x}^\mu$ that fulfil the commutation relations~\cite{dfr1995}
\begin{equation}
\label{eq:xmu_xnu_theta}
[ \hat{x}^\mu, \hat{x}^\nu ] = \mathi \Theta^{\mu\nu} \eqend{,}
\end{equation}
where $\Theta$ is a constant skew-symmetric matrix of the order of the Planck length. While such a quantum spacetime is physically well-motivated (as the solution of the geometrical measurement problem), we feel that nevertheless it is postulated in a rather \emph{ad-hoc} way. Moreover, a constant $\Theta$ matrix is incompatible with Lorentz covariance of the theory, except in two spacetime dimensions where $\Theta^{\mu\nu} \sim \epsilon^{\mu\nu}$ is a Lorentz-invariant choice.\footnote{There are Lorentz-invariant interpretations based on a twisted Poincar\'e symmetry~\cite{chaichiankulishnishijimatureanu2004}; here we refer to the usual untwisted symmetry.}

Describing gravity as the curvature of spacetime, these two approaches should be related, and the question arises how to make a concrete connection. At first sight, it seems as if they describe phenomena at different scales: while perturbative quantum gravity is valid as an effective theory as long as the energy density of the system is well below the density required to create a black-hole, noncommutative geometry on the other hand is introduced to avoid the geometrical measurement problem, i.e., the fact that (classically) a black hole is formed when one tries to localize a spacetime point with arbitrary accuracy, since the energy used for the localization is at some point concentrated in a region smaller than the corresponding Schwarzschild radius. We are thus led to the natural question: are noncommutative geometrical effects already present on lower scales, where perturbative quantum gravity provides a predictive framework? Moreover, is it necessary to introduce noncommutativity by hand, or does it arise through quantum-gravitational effects? We show that the first question can be answered in the affirmative, and the second one in the negative: perturbative quantum gravity in fact \emph{predicts} noncommutativity at the Planck scale, once one clarifies the structure of observables in the quantum theory.

\textbf{Conventions:} We work in four spacetime dimensions, set $\hbar = c = 1$, choose the ``$+++$'' convention of~\cite{mtw} for the metric and curvature tensors, and write $t = x^0$ and $s = y^0$.

\section{Relational observables and perturbative quantum gravity}
\label{sec:relational}

A major obstacle in any theory of quantum gravity is the definition of suitable observables. Since the symmetries of gravity include diffeomorphisms, which move points on the underlying manifold, it is clear that any local field, i.e., any quantity defined at a fixed point is not invariant and can thus not be observable. Perturbative quantum gravity is an effective field theory approach to quantum gravity~\cite{burgess2003}, where one decomposes the full metric $g_{\mu\nu}$ into a background $g^0_{\mu\nu}$ and a perturbation $h_{\mu\nu}$ according to
\begin{equation}
\label{eq:relational_metricdecomp}
g_{\mu\nu} = g^0_{\mu\nu} + \kappa h_{\mu\nu}
\end{equation}
with $\kappa = \sqrt{16 \pi G_\text{N}}$ and quantizes the perturbation on that background. Since the background is fixed, one also restricts to small diffeomorphisms $x^\mu \to x^\mu + \delta_\xi x^\mu = x^\mu - \kappa \xi^\mu$ with parameter $\xi^\mu$ that do not change the background. The changes in tensor fields are obtained using the Lie derivative $\mathcal{L}_{\kappa \xi}$, such that inserting the decomposition~\eqref{eq:relational_metricdecomp} into $\delta_\xi g_{\mu\nu} = \mathcal{L}_{\kappa \xi} g_{\mu\nu}$ we obtain
\begin{splitequation}
\label{eq:relational_deltah}
\delta_\xi h_{\mu\nu} &= \xi^\rho \partial_\rho g_{\mu\nu} + g_{\rho\nu} \partial_\mu \xi^\rho + g_{\rho\mu} \partial_\nu \xi^\rho \\
&= \nabla^0_\mu \xi_\nu + \nabla^0_\nu \xi_\mu \\
&\quad+ \kappa \left( \xi^\rho \nabla^0_\rho h_{\mu\nu} + h_{\rho\mu} \nabla^0_\nu \xi^\rho + h_{\rho\nu} \nabla^0_\mu \xi^\rho \right) \eqend{,}
\end{splitequation}
where $\nabla^0$ is the covariant derivative associated to the Levi-Civita connection of the background metric $g^0_{\mu\nu}$, and we lower and raise all indices with the background metric. At linear order, we recover the well-known result of linearized gravity $\delta^0_\xi h_{\mu\nu} = \mathcal{L}_\xi g^0_{\mu\nu} = \nabla^0_\mu \xi_\nu + \nabla^0_\nu \xi_\mu$, but at higher orders we have the further contribution $\delta^1_\xi h_{\mu\nu} = \mathcal{L}_\xi h_{\mu\nu} = \xi^\rho \nabla^0_\rho h_{\mu\nu} + h_{\rho\mu} \nabla^0_\nu \xi^\rho + h_{\rho\nu} \nabla^0_\mu \xi^\rho$. Consider now a general tensor field $T = T^0 + \kappa T^1 + \bigo{\kappa^2}$, whose linear transformation is given by $\delta^0_\xi T^1 = \mathcal{L}_\xi T^0$. By the Stewart--Walker lemma~\cite{stewartwalker1974}, we have $\delta^0_\xi T^1 = 0$ (such that $T^1$ is an observable in the linearized theory) if and only if $T^0$ is a sum of constant coefficients times a product of delta functions $\delta_\mu^\nu$. This is indeed the case for special backgrounds, such as Minkowski spacetime where $g^0_{\mu\nu} = \eta_{\mu\nu}$ and the background Riemann tensor vanishes $R^0_{\mu\nu\rho\sigma} = 0$, such that the linearized Riemann tensor $R^1_{\mu\nu\rho\sigma} = \partial_\nu \partial_{[\rho} h_{\sigma]\mu} - \partial_\mu \partial_{[\rho} h_{\sigma]\nu}$ is invariant at linear order: $\delta^0_\xi R^1_{\mu\nu\rho\sigma} = 0$. It can further be shown that the linearized Riemann tensor constitutes a complete set of observables, in the sense that any invariant local observable in the linear theory that involves (only) $h_{\mu\nu}$ can be expressed using the linearized Riemann tensor~\cite{higuchi2012}. Also for the de Sitter spacetime, cosmological Friedmann--Lema{\^i}tre--Robertson--Walker (FLRW) spacetimes and black holes one can construct such complete sets of local observables in the linearized theory~\cite{higuchi2012,froebhackhiguchi2017,froebhackkhavkine2018,aksteineranderssonbaeckdahl2019,aksteineretal2021}, and a systematic determination of such sets is possible using a so-called IDEAL characterization of the background spacetime~\cite{ferrandosaez2009,ferrandosaez2010,canepadappiaggikhavkine2018,khavkine2019}.

However, already at second order this breaks down: $T^1 + T^2$ is a (local) observable to second order if and only if $\delta^1_\xi T^1 + \delta^0_\xi T^2 = 0$, but if $T^1$ is nontrivial (i.e., dynamical) this can never be the case. Observables in (perturbative) quantum gravity are thus generically nonlocal, and one possible way to construct them is relationally. Generally, a relational observable is given by the value of one dynamical field of the theory with respect to or in relation to another dynamical field, i.e., at the point where the other field has a certain value. In practice, this means that one chooses four scalar field that serve as dynamical or field-dependent coordinates, and invariant relational observables are given by fields evaluated in this coordinate system. While relational observables have a long history (see~\cite{tambornino2012,goellerhoehnkirklin2022} for reviews), there are some issues that have been solved only recently. In particular, for highly-symmetric backgrounds such as Minkowski spacetime, it is not clear how to define the scalars that make up the dynamical coordinate system in such a way that points on the background can be discriminated, since all curvature scalars --- which otherwise could have been used for this purpose --- vanish on the background. It is of course possible to add scalar fields by hand, but this obviously changes the dynamics~\cite{gieselheroldlisingh2020}. Only recently~\cite{brunettietal2016}, a systematic solution to this problem for cosmological background spacetimes was proposed, in which the coordinates are constructed to all orders in perturbation theory from the gauge-dependent parts of the metric perturbation $h_{\mu\nu}$. Later on~\cite{froeb2018,froeblima2018} this was generalized to other background spacetimes (including Minkowski), and it was shown how to improve the construction in such a way that causality is ensured.

In the approach of~\cite{brunettietal2016,froeb2018,froeblima2018}, the required field-dependent coordinates $X^{(\mu)}$ are determined as the solutions of a scalar differential equation (depending on the full perturbed metric $g_{\mu\nu}$), which on the background reduce to the background coordinates $x^\mu$. For a Minkowski background, this can be taken as~\cite{froeb2018}
\begin{equation}
\label{eq:relational_nabla2x}
\nabla^2 X^{(\mu)} = 0 \eqend{,}
\end{equation}
with the covariant derivative $\nabla$ associated to the Levi-Civita connection of the full metric $g_{\mu\nu}$, since on the background the equation $\partial^2 x^\mu = 0$ is fulfilled for the Cartesian coordinates $x^\mu$. Expanding Eq.~\eqref{eq:relational_nabla2x} to first order with $X^{(\mu)} = x^\mu + \kappa X_1^{(\mu)} + \bigo{\kappa^2}$, we obtain
\begin{equation}
\label{eq:relational_d2x1}
\partial^2 X_1^{(\mu)} = \eta^{\rho\sigma} \Gamma^\mu_{1\rho\sigma} \eqend{,}
\end{equation}
where
\begin{equation}
\Gamma^\mu_{1\rho\sigma} = \frac{1}{2} \eta^{\mu\nu} \left( \partial_\rho h_{\sigma\nu} + \partial_\sigma h_{\rho\nu} - \partial_\nu h_{\rho\sigma} \right)
\end{equation}
is the first-order perturbation of the Christoffel symbol. To obtain an explicit expression for $X_1$, we solve~\eqref{eq:relational_d2x1} using a Green's function $G$ of the scalar d'Alembertian operator $\partial^2$, such that
\begin{equation}
\label{eq:relational_x1sol}
X_1^{(\mu)}(x) = \int G(x,y) \left[ \partial_\rho h^{\rho\mu}(y) - \frac{1}{2} \partial^\mu h(y) \right] \total^4 y \eqend{,}
\end{equation}
where we recall that indices are raised and lowered with the background Minkowski metric $\eta_{\mu\nu}$, and set $h \equiv h^\rho{}_\rho$. Expanding Eq.~\eqref{eq:relational_nabla2x} to higher orders, one can systematically determine the field-dependent coordinates $X^{(\mu)}$ to all orders in perturbation theory.

Since the field-dependent coordinates are obtained as solutions of a scalar Eq.~\eqref{eq:relational_nabla2x}, under a small diffeomorphism with parameter $\kappa \xi^\mu$ they transform as scalars (which is also the reason why we enclosed the index $\mu$ in parentheses). This can be explicitly checked: using the first-order transformation
\begin{equation}
\label{eq:relational_deltaxi_h}
\delta^0_\xi h_{\mu\nu} = \partial_\mu \xi_\nu + \partial_\nu \xi_\mu
\end{equation}
of the metric perturbation, we obtain from Eq.~\eqref{eq:relational_x1sol} that
\begin{splitequation}
\label{eq:relational_deltaxi_x1}
\delta^0_\xi X_1^{(\mu)}(x) &= \int G(x,y) \left[ \partial_\rho \delta^0_\xi h^{\rho\mu}(y) - \frac{1}{2} \partial^\mu \delta^0_\xi h(y) \right] \total^4 y \\
&= \int G(x,y) \partial^2 \xi^\mu(y) \total^4 y \\
&= \xi^\mu(x) = \xi^\rho(x) \partial_\rho x^\mu \eqend{,} \raisetag{2em}
\end{splitequation}
using that the Green's function $G(x,y)$ satisfies $\partial^2 G(x,y) = \delta^4(x-y)$ and integrating by parts for $\xi^\mu$ of compact support. This is exactly the first-order term of the general transformation
\begin{equation}
\delta_\xi X^{(\mu)} = \kappa \xi^\rho \partial_\rho X^{(\mu)} = \mathcal{L}_{\kappa \xi} X^{(\mu)}
\end{equation}
of a scalar quantity, and one can check that this transformation also holds to higher orders. Invariant relational observables can now be constructed to any order in perturbation theory by performing a diffeomorphism from the background coordinate system to the field-dependent coordinate system formed by the $X^{(\mu)}$. For example, from a scalar $S = S^0 + \kappa S^1 + \bigo{\kappa^2}$, we obtain the invariant scalar $\mathcal{S}$ as
\begin{splitequation}
\label{eq:relational_invs}
\mathcal{S}(X) &= S[x(X)] = S^0(X) + \kappa S^1(X) \\
&\quad- \kappa X_1^{(\mu)}(X) \partial_\mu S^0(X) + \bigo{\kappa^2} \eqend{,}
\end{splitequation}
where we used that the inverse of the relation $X^{(\mu)} = x^\mu + \kappa X_1^{(\mu)}(x) + \bigo{\kappa^2}$ reads $x^\mu = X^{(\mu)} - \kappa X_1^{(\mu)}(X) + \bigo{\kappa^2}$ to first order. Using the first-order transformation~\eqref{eq:relational_deltaxi_x1} of $X_1^{(\mu)}$, it is now easy to verify that $\mathcal{S}$ is invariant: using that $S$ transforms according to
\begin{equation}
\delta_\xi S = \mathcal{L}_{\kappa \xi} S = \kappa \xi^\rho \partial_\rho S \eqend{,}
\end{equation}
such that at first order in perturbation theory we have
\begin{equation}
\delta^0_\xi S^1 = \xi^\rho \partial_\rho S^0
\end{equation}
(and the background value $S^0$ is invariant), it follows that
\begin{equation}
\delta^0_\xi \mathcal{S} = \xi^\rho \partial_\rho S^0 - \delta^0_\xi X_1^{(\mu)} \partial_\mu S^0 = 0 \eqend{.}
\end{equation}

These invariant observables have been used to compute various effects in perturbative quantum gravity, such as quantum corrections to the Newtonian gravitational potential~\cite{froebreinverch2022} or to the expansion rate of the early universe~\cite{froeb2018b,lima2021}. However, in this work, we are interested in the field-dependent coordinates $X^{(\mu)}$ themselves. Namely, in the effective field theory that perturbative quantum gravity constitutes, we quantize the metric fluctuations $h_{\mu\nu}$ in the Minkowski background. Since they therefore have a nontrivial commutator, also the coordinates $X^{(\mu)}$ that depend on $h_{\mu\nu}$ will have a nontrivial commutator. We will compute the leading contributions to this induced noncommutativity in the next section.

\section{Noncommutativity of the coordinates}
\label{sec:nc}

To quantize the metric perturbations, we expand the well-known Einstein--Hilbert action $S_\text{EH} = \kappa^{-2} \int R \sqrt{-g} \total^4 x$ to second order around the flat Minkowski background, which gives
\begin{equation}
\label{eq:nc_action2}
S_2 = \frac{1}{2} \int h_{\mu\nu} P^{\mu\nu\rho\sigma} h_{\rho\sigma} \total^4 x
\end{equation}
with the symmetric differential operator
\begin{splitequation}
P^{\mu\nu\rho\sigma} &\equiv \frac{1}{2} \eta^{\mu(\rho} \eta^{\sigma)\nu} \partial^2 - \partial^{(\mu} \eta^{\nu)(\rho} \partial^{\sigma)} \\
&\quad+ \frac{1}{2} \left( \eta^{\mu\nu} \partial^\rho \partial^\sigma + \eta^{\rho\sigma} \partial^\mu \partial^\nu \right) - \frac{1}{2} \eta^{\mu\nu} \eta^{\rho\sigma} \partial^2 \eqend{.}
\end{splitequation}
Since the action is invariant under the first-order transformation~\eqref{eq:relational_deltaxi_h} of the metric perturbation $\delta^0_\xi S_2 = 0$, the operator $P^{\mu\nu\rho\sigma}$ has a kernel consisting of all tensor fields of the form $\partial_\mu \xi_\nu + \partial_\nu \xi_\mu$, and is thus not invertible. To obtain an invertible operator and determine the propagator of $h_{\mu\nu}$, we need to add a gauge-fixing term to the action, and we take the standard de Donder gauge term~\cite{capperleibbrandtramonmedrano1973}
\begin{equation}
S_\text{GF} = - \frac{1}{2} \int \eta^{\mu\nu} H_\mu H_\nu \total^4 x
\end{equation}
with
\begin{equation}
\label{eq:nc_hmudef}
H_\mu \equiv \partial^\nu h_{\mu\nu} - \frac{1}{2} \partial_\mu h \eqend{.}
\end{equation}
The sum of the second order Einstein--Hilbert action $S_2$ and the gauge-fixing action $S_\text{GF}$ can again be written in the form~\eqref{eq:nc_action2} with the new symmetric differential operator
\begin{equation}
\tilde{P}^{\mu\nu\rho\sigma} \equiv \frac{1}{2} \eta^{\mu(\rho} \eta^{\sigma)\nu} \partial^2 - \frac{1}{4} \eta^{\mu\nu} \eta^{\rho\sigma} \partial^2 \eqend{,}
\end{equation}
which is now invertible. The graviton propagator, the expectation value of the time-ordered product of two metric perturbations $h_{\mu\nu}$
\begin{equation}
G^\text{F}_{\mu\nu\rho\sigma}(x,x') \equiv - \mathi \expect{ \mathcal{T} h_{\mu\nu}(x) h_{\rho\sigma}(x') } \eqend{,}
\end{equation}
is the fundamental solution with Feynman boundary conditions,
\begin{equation}
\label{eq:nc_pg_delta}
\tilde{P}^{\alpha\beta\mu\nu} G^\text{F}_{\mu\nu\rho\sigma}(x,x') = \delta^\alpha_{(\rho} \delta^\beta_{\sigma)} \delta^4(x-x') \eqend{.}
\end{equation}
One easily verifies that~\cite{radkowski1970}
\begin{equation}
\label{eq:nc_gfmunurhosigma}
G^\text{F}_{\mu\nu\rho\sigma}(x,x') = \left( 2 \eta_{\mu(\rho} \eta_{\sigma)\nu} - \eta_{\mu\nu} \eta_{\rho\sigma} \right) G^\text{F}(x,x')
\end{equation}
with the massless scalar Feynman propagator~\cite{peskinschroeder}
\begin{equations}[eq:nc_gf]
\begin{split}
G^\text{F}(x,x') &= - \int \frac{\mathe^{\mathi p (x-x')}}{p^2 - \mathi 0} \frac{\total^4 p}{(2\pi)^4} \\
&= \int \tilde{G}^\text{F}(\vec{p},t,t') \, \mathe^{\mathi \vec{p} (\vec{x}-\vec{x}')} \frac{\total^3 \vec{p}}{(2\pi)^3} \eqend{,}
\end{split} \\
\tilde{G}^\text{F}(\vec{p},t,t') &= - \mathi \frac{\mathe^{- \mathi \abs{\vec{p}} \abs{t-t'}}}{2 \abs{\vec{p}}} \eqend{,}
\end{equations}
is a solution of~\eqref{eq:nc_pg_delta} with the required Feynman boundary conditions.

Since the action $S_2 + S_\text{GF}$ is quadratic, we are dealing with a theory of free fields, such that the commutator of two metric perturbations $h_{\mu\nu}$ is proportional to the identity $\1$. Because the first-order correction $X_1^{(\mu)}$~\eqref{eq:relational_x1sol} is linear in $h_{\mu\nu}$, also the commutator of two $X_1^{(\mu)}$ is proportional to the identity, and we can compute it by computing its expectation value:
\begin{splitequation}
\label{eq:nc_x1comm}
&\left[ X_1^{(\mu)}(x), X_1^{(\nu)}(x') \right] \\
&= \expect{ X_1^{(\mu)}(x) X_1^{(\nu)}(x') - X_1^{(\nu)}(x') X_1^{(\mu)}(x) } \1 \eqend{.}
\end{splitequation}
However, if we only consider the Feynman propagator $G^\text{F}$, all we can compute are time-ordered correlation functions (and from these $S$-matrix elements). To compute true expectation values, we also need the (positive and negative frequency) Wightman functions $G^+$ and $G^-$ and the anti-time-ordered (or Dyson) propagator $G^\text{D}$, all of which are simple modifications of~\eqref{eq:nc_gf}, which only differ in the time dependence in the exponential,
\begin{equations}[eq:nc_gpmd]
\tilde{G}^+(\vec{p},t,t') &= - \mathi \frac{\mathe^{- \mathi \abs{\vec{p}} (t-t')}}{2 \abs{\vec{p}}} \eqend{,} \\
\tilde{G}^-(\vec{p},t,t') &= - \mathi \frac{\mathe^{\mathi \abs{\vec{p}} (t-t')}}{2 \abs{\vec{p}}} \eqend{,} \\
\tilde{G}^\text{D}(\vec{p},t,t') &= - \mathi \frac{\mathe^{\mathi \abs{\vec{p}} \abs{t-t'}}}{2 \abs{\vec{p}}} \eqend{.}
\end{equations}
These different two-point functions can be unified in the so-called in-in, Schwinger--Keldysh or closed-time-path formalism (see e.g.~\cite{chousuhaolu1985,jordan1986}), where one replaces the time integration from $-\infty$ to $+\infty$ and the time-ordering of fields (the usual in-out formalism) by a path going from $-\infty$ to $+\infty$ and back to $-\infty$ and the path-ordering of fields along this integration contour. Denoting fields on the forward part of the contour with a ``$+$'' and fields on the backward part of the contour with a ``$-$'', the path-ordered two point function
\begin{equation}
\label{eq:nc_gab_munurhosigma}
G^{AB}_{\mu\nu\rho\sigma}(x,x') \equiv - \mathi \expect{ \mathcal{P} h^A_{\mu\nu}(x) h^B_{\rho\sigma}(x') }
\end{equation}
with $A,B = \pm$ can then take four possible values:
\begin{equations}
\begin{split}
G^{++}_{\mu\nu\rho\sigma}(x,x') &= - \mathi \expect{ \mathcal{T} h^+_{\mu\nu}(x) h^+_{\rho\sigma}(x') } \\
&= G^\text{F}_{\mu\nu\rho\sigma}(x,x') \eqend{,}
\end{split} \\
\begin{split}
G^{+-}_{\mu\nu\rho\sigma}(x,x') &= - \mathi \expect{ h^-_{\rho\sigma}(x') h^+_{\mu\nu}(x) } \\
&= G^-_{\mu\nu\rho\sigma}(x,x') \eqend{,}
\end{split} \\
\begin{split}
G^{-+}_{\mu\nu\rho\sigma}(x,x') &= - \mathi \expect{ h^-_{\mu\nu}(x) h^+_{\rho\sigma}(x') } \\
&= G^+_{\mu\nu\rho\sigma}(x,x') \eqend{,}
\end{split} \\
\begin{split}
G^{--}_{\mu\nu\rho\sigma}(x,x') &= - \mathi \expect{ \overline{\mathcal{T}} h^-_{\mu\nu}(x) h^-_{\rho\sigma}(x') } \\
&= G^\text{D}_{\mu\nu\rho\sigma}(x,x') \eqend{,}
\end{split}
\end{equations}
since fields on the backward part of the contour are always ``later'' than fields on the forward part.

It follows that the field-dependent coordinates $X^{(\mu)}$ can also either be on the forward or the backward part of the contour, and furthermore the time integration contour in the first-order correction $X_1^{(\mu)}$~\eqref{eq:relational_x1sol} must be the full path running from $-\infty$ to $+\infty$ and back again, with the appropriate Green's function or propagator. On the forward contour, this is
\begin{splitequation}
\label{eq:nc_x1p}
X_1^{+(\mu)}(x) &= \int G^\text{F}(x,y) H_+^\mu(y) \total^4 y \\
&\quad- \int G^-(x,y) H_-^\mu(y) \total^4 y \eqend{,}
\end{splitequation}
while on the backward contour we have
\begin{splitequation}
\label{eq:nc_x1m}
X_1^{-(\mu)}(x) &= \int G^+(x,y) H_+^\mu(y) \total^4 y \\
&\quad- \int G^\text{D}(x,y) H_-^\mu(y) \total^4 y \eqend{,}
\end{splitequation}
where to shorten the expressions we defined
\begin{equation}
\label{eq:nc_hamu_def}
H_A^\mu \equiv \partial_\rho h_A^{\rho\mu} - \frac{1}{2} \partial^\mu h_A \eqend{.}
\end{equation}
Note that in both cases we have split the contour into forward and backward parts, and for the backward part where the time integration runs originally from $+\infty$ to $-\infty$ have switched integration limits, which resulted in a minus sign. Since the Wightman functions are solutions of the (massless) Klein--Gordon equation $\partial^2 G^+(x,y) = 0 = \partial^2 G^-(x,y)$, while the Feynman and Dyson propagators are fundamental solutions $\partial^2 G^\text{F}(x,y) = \delta^4(x-y) = - \partial^2 G^\text{D}(x,y)$ as can easily be checked from the explicit expressions~\eqref{eq:nc_gf} and~\eqref{eq:nc_gpmd}, we still have
\begin{equation}
\partial^2 X_1^{A(\mu)} = H_A^\mu = \eta^{\rho\sigma} \Gamma^{A\mu}_{1\rho\sigma} \eqend{,}
\end{equation}
such that the defining relation~\eqref{eq:relational_d2x1} for the first-order coordinate corrections still holds on both parts of the contour.

The commutator of two $X_1^{(\mu)}$~\eqref{eq:nc_x1comm} is then obtained as
\begin{splitequation}
\label{eq:nc_x1_expect}
&\left[ X_1^{(\mu)}(x), X_1^{(\nu)}(x') \right] \\
&= \int \left[ F^{\mu\nu}_{-+}(t,t',\vec{p}) - F^{\mu\nu}_{+-}(t,t',\vec{p}) \right] \mathe^{\mathi \vec{p} (\vec{x}-\vec{x}')} \frac{\total^3 \vec{p}}{(2\pi)^3} \1 \eqend{,}
\end{splitequation}
where we defined $F^{\mu\nu}_{AB}$ as the Fourier transform of the path-ordered expectation value
\begin{splitequation}
\label{eq:nc_fabmunu_def}
&\expect{ \mathcal{P} X_1^{A(\mu)}(x) X_1^{B(\nu)}(x') } \\
&= \int F^{\mu\nu}_{AB}(t,t',\vec{p}) \, \mathe^{\mathi \vec{p} (\vec{x}-\vec{x}')} \frac{\total^3 \vec{p}}{(2\pi)^3} \eqend{.}
\end{splitequation}
Inserting the expressions~\eqref{eq:nc_x1p} and~\eqref{eq:nc_x1m} for the $X_1^{A(\mu)}$ and passing to Fourier space for all spatial variables, we obtain
\begin{widetext}
\begin{splitequation}
\label{eq:nc_fpmmunu}
&F^{\mu\nu}_{+-}(t,t',\vec{p}) = \mathi \eta^{\mu\nu} \iint \bigg[ \tilde{G}^{++}(t,s,\vec{p}) \Big[ \tilde{G}^{-+}(t',s',\vec{p}) \left( \partial_s^2 + \vec{p}^2 \right) \tilde{G}^{++}(s,s',\vec{p}) - \tilde{G}^{--}(t',s',\vec{p}) \left( \partial_s^2 + \vec{p}^2 \right) \tilde{G}^{+-}(s,s',\vec{p}) \Big] \\
&\qquad- \tilde{G}^{+-}(t,s,\vec{p}) \Big[ \tilde{G}^{-+}(t',s',\vec{p}) \left( \partial_s^2 + \vec{p}^2 \right) \tilde{G}^{-+}(s,s',\vec{p}) - \tilde{G}^{--}(t',s',\vec{p}) \left( \partial_s^2 + \vec{p}^2 \right) \tilde{G}^{--}(s,s',\vec{p}) \Big] \bigg] \total s \total s' \eqend{,} \raisetag{1.9em}
\end{splitequation}
\end{widetext}
where we also used that the various propagators/two-point functions~\eqref{eq:nc_gf} and~\eqref{eq:nc_gpmd} only depend on the time difference $t-t'$ to convert derivatives with respect to $t'$ into derivatives with respect to $t$, as well as
\begin{equation}
\label{eq:nc_fpm_fmp}
F^{\mu\nu}_{+-}(t,t',\vec{p}) = F^{\mu\nu}_{-+}(t',t,\vec{p}) \eqend{.}
\end{equation}
Since the Wightman functions are solutions of the Klein--Gordon equation, we have $\left( \partial_s^2 + \vec{p}^2 \right) \tilde{G}^{-+}(s,s',\vec{p}) = \left( \partial_s^2 + \vec{p}^2 \right) \tilde{G}^{+-}(s,s',\vec{p}) = 0$, as one can also check directly from the explicit expressions~\eqref{eq:nc_gpmd}. On the other hand, the Feynman and Dyson propagators are fundamental solutions with $\left( \partial_s^2 + \vec{p}^2 \right) \tilde{G}^{++}(s,s',\vec{p}) = - \left( \partial_s^2 + \vec{p}^2 \right) \tilde{G}^{--}(s,s',\vec{p}) = - \delta(s-s')$ (which can be checked from~\eqref{eq:nc_gf} and~\eqref{eq:nc_gpmd}). Therefore, the middle two terms in~\eqref{eq:nc_fpmmunu} vanish and for the other two we can use the $\delta$ to perform the integral over $s'$ and obtain
\begin{splitequation}
\label{eq:nc_fpmmunu_2}
F^{\mu\nu}_{+-}(t,t',\vec{p}) &= - \mathi \eta^{\mu\nu} \int \Bigg[ \tilde{G}^{++}(t,s,\vec{p}) \tilde{G}^{-+}(t',s,\vec{p}) \\
&\qquad- \tilde{G}^{+-}(t,s,\vec{p}) \tilde{G}^{--}(t',s,\vec{p}) \Bigg] \total s \eqend{.}
\end{splitequation}
If we now naively try to evaluate this integral, we obtain a divergence as $s \to - \infty$. This infrared divergence arises because the massless propagator only decays slowly, with a power of the distance between the two points. Let us for a moment consider classical metric perturbations $h_{\mu\nu}$, i.e., we make no difference between the ``$+$'' fields on the forward part of the contour and the ``$-$'' fields on the backward part. Then the expressions for both $X_1^{+(\mu)}$~\eqref{eq:nc_x1p} and $X_1^{-(\mu)}$~\eqref{eq:nc_x1m} reduce to
\begin{equation}
\label{eq:nc_x1mu_gret}
X_1^{(\mu)}(x) = \int G^\text{ret}(x,y) H^\mu(y) \total^4 y
\end{equation}
with the retarded propagator
\begin{equation}
G^\text{ret}(x,x') \equiv \Theta(t-t') \left[ G^+(x,x') - G^-(x,x') \right] \eqend{.}
\end{equation}
Since $G^+(x,x') = G^-(x,x')$ if $x$ and $x'$ are spacelike separated, the retarded propagator vanishes outside the light cone, such that the integration in~\eqref{eq:nc_x1mu_gret} is restricted to the past light cone of the point $x$. For each fixed time, the spatial integration is therefore finite (which justifies the use of the spatial Fourier transform in the above computations), but as $s \to - \infty$ the volume of the spatial integration region grows. To obtain a finite result, we add a convergence factor $\exp(\epsilon \abs{\vec{p}} s)$ with $\epsilon > 0$ to the two-point function~\eqref{eq:nc_fpmmunu_2}, and take the limit $\epsilon \to 0$ after integration. The addition of this convergence factor can also be interpreted as a slight deformation of the time integration contour into the complex plane $s \to s (1 \pm \mathi \epsilon)$ (depending on the exact exponential factor), which selects the full interacting vacuum state of the theory~\cite{peskinschroeder,froebrouraverdaguer2012,baumgartsundrum2021}. Even though we consider a theory of free fields, there are nontrivial interactions because of the integration over the past light cone; the convergence factor can thus also be interpreted as an adiabatic cutoff of the interaction~\cite{lippmannschwinger1950}.

\begin{widetext}
Inserting the explicit expressions for the propagators~\eqref{eq:nc_gf} and~\eqref{eq:nc_gpmd} into the two-point function~\eqref{eq:nc_fpmmunu_2}, we therefore obtain
\begin{splitequation}
\label{eq:nc_fpmmunu_3}
F^{\mu\nu}_{+-}(t,t',\vec{p}) &= \frac{\mathi \eta^{\mu\nu}}{4 \vec{p}^2} \lim_{\epsilon \to 0} \int \mathe^{\epsilon \abs{\vec{p}} s} \Bigg[ \mathe^{- \mathi \abs{\vec{p}} \abs{t-s} - \mathi \abs{\vec{p}} (t'-s)} - \mathe^{\mathi \abs{\vec{p}} (t-s) + \mathi \abs{\vec{p}} \abs{t'-s}} \Bigg] \total s \\
&= \frac{\mathi \eta^{\mu\nu}}{4 \vec{p}^2} \lim_{\epsilon \to 0} \Bigg[ \mathe^{- \mathi \abs{\vec{p}} (t+t')} \frac{\mathe^{(2 \mathi + \epsilon) \abs{\vec{p}} t}}{(2 \mathi + \epsilon) \abs{\vec{p}}} - \mathe^{\mathi \abs{\vec{p}} (t+t')} \frac{\mathe^{(- 2 \mathi + \epsilon) \abs{\vec{p}} t'}}{(- 2 \mathi + \epsilon) \abs{\vec{p}}} + \mathe^{\mathi \abs{\vec{p}} (t-t')} \frac{\mathe^{\epsilon \abs{\vec{p}} t'} - \mathe^{\epsilon \abs{\vec{p}} t}}{\epsilon \abs{\vec{p}}} \Bigg] \\
&= \frac{\eta^{\mu\nu}}{4 \abs{\vec{p}}^3} \mathe^{\mathi \abs{\vec{p}} (t-t')} \Big[ 1 - \mathi \abs{\vec{p}} (t-t') \Big] \eqend{,}
\end{splitequation}
which is finite as required. Using also the relation~\eqref{eq:nc_fpm_fmp}, the commutator~\eqref{eq:nc_x1_expect} thus reads
\begin{splitequation}
\left[ X_1^{(\mu)}(x), X_1^{(\nu)}(x') \right] = \int \frac{\mathi \eta^{\mu\nu}}{2 \abs{\vec{p}}^3} \Big[ \cos\left[ \abs{\vec{p}} (t-t') \right] \abs{\vec{p}} (t-t') - \sin\left[ \abs{\vec{p}} (t-t') \right] \Big] \mathe^{\mathi \vec{p} (\vec{x}-\vec{x}')} \frac{\total^3 \vec{p}}{(2\pi)^3} \1 \eqend{,}
\end{splitequation}
and it only remains to perform the inverse Fourier transform. Choosing spherical coordinates such that $\vec{p} (\vec{x}-\vec{x}') = \abs{\vec{p}} r \cos \theta$ with $r \equiv \abs{\vec{x}-\vec{x}'}$, we obtain
\begin{splitequation}
\left[ X_1^{(\mu)}(x), X_1^{(\nu)}(x') \right] &= \frac{\mathi \eta^{\mu\nu}}{2 (2\pi)^2} \int_0^\infty \int_0^\pi \Big[ \cos\left[ \abs{\vec{p}} (t-t') \right] (t-t') - \abs{\vec{p}}^{-1} \sin\left[ \abs{\vec{p}} (t-t') \right] \Big] \mathe^{\mathi \abs{\vec{p}} r \cos \theta} \sin \theta \total \theta \total \abs{\vec{p}} \, \1 \\
&= \frac{\mathi \eta^{\mu\nu}}{(2\pi)^2} \int_0^\infty \Big[ \cos\left[ \abs{\vec{p}} (t-t') \right] (t-t') - \abs{\vec{p}}^{-1} \sin\left[ \abs{\vec{p}} (t-t') \right] \Big] \frac{\sin( \abs{\vec{p}} r )}{\abs{\vec{p}} r} \total \abs{\vec{p}} \, \1 \eqend{.}
\end{splitequation}
For the integral over $\abs{\vec{p}}$, we use the exponential integral~\cite[Eqs.~(6.6.2) and~(6.7.1)]{dlmf}
\begin{equation}
\int_0^\infty \frac{\mathe^{- a t}}{t + b} \total t = \mathe^{a b} E_1(a b) = - \gamma - \ln(a b) + \bigo{b} \eqend{,}
\end{equation}
valid for $\Re a, \Re b > 0$, as well as the corollary
\begin{equation}
\int_0^\infty \frac{\mathe^{- a t}}{(t + b)^2} \total t = \int_0^\infty \left( - a - \partial_t \right) \frac{\mathe^{- a t}}{t + b} \total t = \frac{1}{b} - a \, \mathe^{a b} E_1(a b) \eqend{.}
\end{equation}
This results in
\begin{splitequation}
\left[ X_1^{(\mu)}(x), X_1^{(\nu)}(x') \right] &= - \frac{\eta^{\mu\nu}}{(4 \pi)^2 r} \lim_{\delta,\epsilon \to 0} \int_0^\infty \mathe^{- \delta \abs{\vec{p}}} \bigg[ \frac{(t-t')}{\abs{\vec{p}} + \epsilon} \Big[ \mathe^{\mathi \abs{\vec{p}} (t-t'-r)} - \mathe^{\mathi \abs{\vec{p}} (t-t'+r)} - \mathe^{- \mathi \abs{\vec{p}} (t-t'-r)} + \mathe^{- \mathi \abs{\vec{p}} (t-t'+r)} \Big] \\
&\hspace{7em}+ \frac{\mathi}{(\abs{\vec{p}} + \epsilon)^2} \Big[ \mathe^{\mathi \abs{\vec{p}} (t-t'-r)} - \mathe^{\mathi \abs{\vec{p}} (t-t'+r)} - \mathe^{- \mathi \abs{\vec{p}} (t-t'+r)} + \mathe^{- \mathi \abs{\vec{p}} (t-t'-r)} \Big] \bigg] \total \abs{\vec{p}} \, \1 \\
&= \frac{\eta^{\mu\nu}}{(4 \pi)^2 r} \lim_{\delta \to 0} \int_0^\infty \bigg[ (t-t') \Big[ \ln[ \delta - \mathi (t-t'-r) ] - \ln[ \delta + \mathi (t-t'-r) ] + \ln[ \delta + \mathi (t-t'+r) ] \\
&\hspace{7em}- \ln[ \delta - \mathi (t-t'+r) ] \Big] - (t-t'-r) \ln(\delta - \mathi (t-t'-r)) \\
&\hspace{7em}+ (t-t'+r) \ln(\delta - \mathi (t-t'+r)) - (t-t'+r) \ln(\delta + \mathi (t-t'+r)) \\
&\hspace{7em}+ (t-t'-r) \ln(\delta + \mathi (t-t'-r)) \bigg] \total \abs{\vec{p}} \, \1 \eqend{,} \raisetag{2em}
\end{splitequation}
and using the well-known
\begin{equation}
\lim_{\delta \to 0} \ln(\delta + \mathi a) = \ln\abs{a} + \frac{\mathi \pi}{2} \sgn(a) \eqend{,}
\end{equation}
it follows that
\begin{splitequation}
\left[ X_1^{(\mu)}(x), X_1^{(\nu)}(x') \right] &= - \mathi \frac{\eta^{\mu\nu}}{16 \pi} \left[ \sgn(t-t'+r) + \sgn(t-t'-r) \right] \1 \\
&= - \mathi \frac{\eta^{\mu\nu}}{16 \pi} \sgn(t-t') \left[ 1 + \sgn(\abs{t-t'}-r) \right] \1 \eqend{.}
\end{splitequation}
The right-hand side vanishes for spacelike separations where $r > \abs{t-t'}$, such that (with $(x-x')^2 = r^2 - (t-t')^2$) we can rewrite the result as
\begin{equation}
\label{eq:nc_x1comm_result}
\left[ X_1^{(\mu)}(x), X_1^{(\nu)}(x') \right] = - \mathi \frac{\eta^{\mu\nu}}{8 \pi} \sgn(t-t') \Theta[ -(x-x')^2 ] \1 \eqend{.}
\end{equation}
\end{widetext}
That is, the commutator of two $X_1^{(\mu)}$ is vanishing when $x$ and $x'$ are spacelike separated, and nonvanishing only for timelike separations, with the sign depending on whether the second $X_1^{(\nu)}$ is in the future or in the past of the first $X_1^{(\mu)}$. Moreover, it is clear that the result is fully Lorentz invariant.

At this point a crucial difference to other approaches to noncommutative geometry becomes clear. In conventional approaches, the noncommutative coordinates are operators in some abstract space, and the classical geometry only emerges from their spectrum. In perturbative quantum gravity, there exist instead physical events (points) $p$ which are defined relationally, e.g., the 1994 coincidence of comet Shoemaker-Levy 9 with Jupiter. On the background, these events are described by commuting coordinates $x^\mu = x^\mu(p)$, while in the full theory they are described by the noncommuting coordinates $X^{(\mu)} = X^{(\mu)}(p)$ [which inherit the noncommutativity from the metric perturbations~\eqref{eq:relational_x1sol}]. Perturbatively, we have $X^{(\mu)}(p) = x^\mu(p) + \kappa X_1^{(\mu)}(p) + \ldots$, and since the background coordinates discriminate the points $p$, we can also write $X_1^{(\mu)}(x)$ as in~\eqref{eq:nc_x1comm_result}. That is, the spacelike (or timelike) separation of $x$ and $x'$ actually refers to the spacelike (or timelike) separation of the physical events which on the background are described by $x$ and $x'$.

For the commutator of two $X^{(\mu)}$, we then use that the background coordinates $x^\mu$ commute with everything, such that
\begin{splitequation}
\label{eq:nc_xcomm_result}
&\left[ X^{(\mu)}, Y^{(\nu)} \right] = \kappa^2 \left[ X_1^{(\mu)}(x), X_1^{(\nu)}(y) \right] + \bigo{\kappa^3} \\
&= - \mathi \kappa^2 \frac{\eta^{\mu\nu}}{8 \pi} \sgn(X^0-Y^0) \Theta[ -(X-Y)^2 ] + \bigo{\kappa^3} \eqend{,} \raisetag{4em}
\end{splitequation}
where $X^{(\mu)} = X^{(\mu)}(p)$ and $Y^{(\mu)} = Y^{(\mu)}(q)$ refer to two distinct physical events $p$ and $q$. Using that $\kappa^2 = 16 \pi G_\text{N} = 16 \pi \ell_\text{Pl}^2$, we see that the result is proportional to the squared Planck length $\ell_\text{Pl}$, which here appears naturally. The most general form of the commutator is given by
\begin{equation}
[ X^\mu, Y^\nu ] = \mathi \Theta^{\mu\nu}(X,Y) \eqend{,}
\end{equation}
where to ensure the antisymmetry of the commutator we must have
\begin{equation}
\label{eq:nc_comm_antisym}
\Theta^{\mu\nu}(X,Y) = - \Theta^{\nu\mu}(Y,X) \eqend{,}
\end{equation}
and to ensure its reality (since the $X^\mu$ should correspond to Hermitean operators) $[ \Theta^{\mu\nu}(X,Y) ]^\dagger = \Theta^{\mu\nu}(X,Y)$. In the postulated realization~\eqref{eq:xmu_xnu_theta}, the approach of~\cite{dfr1995}, $\Theta^{\mu\nu}$ is a constant and thus necessarily must be antisymmetric in its indices $\Theta^{\nu\mu} = - \Theta^{\mu\nu}$ to fulfill the condition~\eqref{eq:nc_comm_antisym}. In contrast, for the result presented here $\Theta^{\mu\nu}(X,Y) = - 2 \ell_\text{Pl}^2 \eta^{\mu\nu} \sgn(X^0-Y^0) \Theta[ -(X-Y)^2 ]$ depends on the coordinates $X$ and $Y$, and so can be symmetric in its indices, since the antisymmetry~\eqref{eq:nc_comm_antisym} comes from the sign $\sgn(X^0-Y^0)$.

\section{Discussion}
\label{sec:discussion}

We have shown that noncommutative geometry is an intrinsic part of perturbative quantum gravity. Rather than being postulated \emph{ad hoc}, it arises naturally when one clarifies the structure of observables in the quantum theory. Namely, physical observables which are invariant under (small) diffeomorphisms can be constructed in a relational way, by evaluating the quantity of interest in a field-dependent coordinate system. This coordinate system is constructed from the gauge-dependent parts of the metric perturbation, in such a way that the resulting observable is gauge independent. Since the quantized metric perturbation has a nontrivial commutator, it follows that also the field-dependent coordinates do not commute, with the induced noncommutativity to leading order given by~\eqref{eq:nc_xcomm_result}.

It is clear that this emerging noncommutative structure has various advantages over conventional approaches: first, it is induced and represented by elements of the theory under consideration, and it is not necessary to introduce any new elements \emph{ad hoc}. In particular, perturbative quantum gravity is a very conservative approach that results from the application of the well-established methods of Lagrangian quantum field theory to the Einstein--Hilbert Lagrangian describing classical gravity. We thus expect that any theory of quantum gravity has to reproduce the predictions of perturbative quantum gravity in its region of validity, which includes this emerging noncommutative structure. Second, the induced noncommutativity~\eqref{eq:nc_xcomm_result} is compatible with microcausality: it vanishes for spacelike separations and is constant within the light cone, with the sign depending on which of the two coordinates is in the future. In particular, it is manifestly Poincar\'e invariant without the need to twist the symmetry. Third, the length scale where the noncommutativity becomes important also arises naturally from the theory, and doesn't need to be postulated. In our case, this is the Planck length, but there are theories where the natural length scale is much larger than the Planck length, such as braneworld models with warped extra dimensions~\cite{randallsundrum1999a,randallsundrum1999b} or models with a large number of fields~\cite{dvali2010}. In these theories, we would also expect that the scale of the noncommutativity is much larger than the Planck length, and it would be interesting to extend our results to these models.

It would also be interesting to derive a generalized uncertainty principle (GUP)~\cite{scardigli1999,adlersantiago1999,scardiglicasadio2003,jizbakleinertscardigli2010,tawfikdiab2014,casadioscardigli2014,scardiglilambiasevagenas2017} (at the noncommutativity scale) from our result. This would give a sound derivation of a GUP from a conservative approach to quantum gravity, without the need to postulate it or derive it by semiclassical arguments, and which then could be experimentally tested~\cite{dasvagenas2008,hossenfelder2013,scardiglicasadio2015}. However, in contrast to the well-known quantum-mechanical Heisenberg uncertainty principle relating (say) the uncertainties in the measurement of a particle's position and momentum, in quantum gravity it is impossible to repeatedly measure the coordinates $X^{(\mu)}$ of the \emph{same} event $p$. The operational meaning of the standard deviation $\Delta_X$ is therefore not clear, and we leave to future work the question how one can extract observable results from the commutator~\eqref{eq:nc_xcomm_result}. What we can assert so far, and what microcausality ensures, is that any measurement cannot influence other measurements done at spacelike separations, independently of the state in which the system is prepared~\cite{hellwigkraus1970,doplicher2018,fewsterverch2018}. This impossibility of superluminal signalling is reflected in the fundamental commutator~\eqref{eq:nc_xcomm_result}, which is vanishing outside the light cone.

Because we are working within the effective field theory of perturbative quantum gravity, the validity of our results is assured for length scales (well) above the Planck length, such that higher orders can be neglected in comparison to the leading one~\eqref{eq:nc_xcomm_result}. In particular, the causal relation between the events described by the noncommutative coordinates is to leading order the same as the one of the background coordinates~\eqref{eq:nc_x1comm_result}, which are commuting; any noncommutativity would only appear at higher orders. To infer from our result strong statements, for example about the resolution of black hole singularities~\cite{nicolinismailagicspalluci2006}, would unfortunately lie outside the range of validity of the effective field theory. However, they give hints on how a full theory of quantum gravity could naturally incorporate noncommutativity. To extend the validity of our result to smaller distances, one would have to compute higher orders and resum the leading corrections, which is a difficult but not impossible task. In this regard, one important question that needs to be investigated further and clarified is the topology of a noncommutative spacetime, and in particular how causal orderings can be defined for arbitrary (not necessarily small) noncommutativity. There are of course concrete proposals, for example for Riemannian spectral triples~\cite{connes2013}, but the Lorentzian case presents extra challenges~\cite{besnard2009}: even for a classical Lorentzian manifold there can be a mismatch between the topology of the underlying manifold and the causal ordering induced by the Lorentzian metric~\cite{papadopoulos2021}; see~\cite{finstermuchpapadopoulos2021} for an overview of results and open questions. However, before tackling these foundational questions we first want to collect more information in concrete examples, and generalize our results to other interesting background spacetimes such as de Sitter or cosmological spacetimes, and possibly higher orders in perturbation theory. It would also be very interesting to see how the results depend on the choice of field-dependent coordinates~\eqref{eq:relational_nabla2x}, even though generalized harmonic coordinates are a natural choice that also appears in other contexts, for example matrix models~\cite{steinacker2010}.

\hfill

\begin{acknowledgments}
M.B.F. acknowledges the support by the Deutsche Forschungsgemeinschaft (DFG, German Research Foundation) --- Project No. 396692871 within the Emmy Noether Grant No. CA1850/1-1 and Project No. 406116891 within the Research Training Group RTG 2522/1. A.M. acknowledges the support by the DFG within the Sonderforschungsbereich (SPP, Priority Program) 2026 ``Geometry at Infinity''. We thank Rainer Verch for comments, Roberto Casadio and Fabio Scardigli for reference suggestions, and the participants of the 2022 Corfu ``Workshop on Noncommutative and generalized geometry in string theory, gauge theory and related physical models'', in particular Paolo Aschieri, Eugenia Boffo, and Harold Steinacker, as well as Nikolaos Kalogeropoulos and the anonymous referee for discussions, comments, and reference suggestions.
\end{acknowledgments}

\bibliography{literature}
\end{document}